\documentclass[twocolumn,showpacs,amsmath,amssymb,byrevtex,endfloats*,citeautoscript,altaffilletter,a4paper]{revtex4}

\usepackage{graphicx}
\usepackage{amsfonts}
\usepackage{mathrsfs}
\usepackage{amsmath}

\newcommand{\Pop}{{\hat{\rm P}}}

\newcommand{\Hop}{{\hat H}}

\newcommand{\geqsi}{\; {\scriptstyle {> \atop \sim}} \;}

\begin{document}

\title{Multifractal fluctuations in the survival probability of an open quantum system}

\author{Angelo Facchini}
\email{a.facchini@unisi.it}
\affiliation{Center for the Study of Complex Systems, University of Siena, Via Tommaso Pendola 37, I-53100 Siena}

\author{Sandro Wimberger}
\affiliation{CNR-INFM and Dipartimento di Fisica ``Enrico Fermi'',
         Universit\`a degli Studi di Pisa, Largo Pontecorvo 3, I-56127 Pisa
         }
\author{Andrea Tomadin}
\affiliation{Scuola Normale Superiore, Piazza dei Cavalieri 7, I-56126 Pisa}

\begin{abstract}
We predict a multifractal behaviour of transport in the deep
quantum regime for the opened $\delta-$kicked rotor model. Our
analysis focuses on intermediate and large scale correlations
in the transport signal and generalizes previously found
parametric {\em mono}-fractal
fluctuations in the quantum survival probability on small scales.
\end{abstract}


\maketitle

\section{Introduction}
\label{sec:Intro}

Multifractal analysis of fluctuating signals is a widely applied method to
characterize complexity on many scales in classical dynamics \cite{Ott1993},
or in the analysis of a given time series (without any a priori knowledge
on the underlying dynamical system which generated the series) \cite{Kantz}.

On the quantum level, multifractal behaviour was found in the scaling of eigenfunctions
in solid-state transport problems \cite{Qmulti}. As far as we know, there have been, however,
very few attempts to use the method of multifractal analysis to directly characterize transport
properties such as conductance (across a solid state sample) or the survival probability
(in open, decaying systems). Often it is indirectly argued that the multifractal structure
of the wave functions at critical points (at the crossover between the localized and
the extended regime) imprints itself on the scaling of transport coefficients
\cite{SM2005}. Other works found a fractal scaling of {\em local} transport quantities,
such as hopping amplitudes \cite{BKAHB2001} or two-point correlations \cite{JMZ1999}. At
criticality \cite{JMZ1999} predicts, e.g., a multifractal scaling of the two-point
conductance between two small interior probes within the transporting sample.

In this paper, we directly study the fluctuations properties of a
{\em global} conductance like quantity in a regime of {\em strong
localization} (Anderson or dynamical localization in our context
of quantum dynamical systems). The studied quantity is the
survival probability of an open, classically chaotic system, which
in the deep quantum realm was found to obey a monofractal scaling
if certain conditions on the quantum eigenvalue spectrum are
fulfilled \cite{GT2001,TMW2006}. In particular, the distribution
of decay rates of the weakly opened system needs to obey a
power-law with an exponent $\gamma\sim -1$, which translates into
an analytic prediction for the corresponding box counting
dimension (of the survival probability as a function of a proper
scan parameter): $D_{\rm BC} \simeq 2 - |\gamma|/2$.

A more detailed, yet preliminary numerical analysis of the decay rate
distribution for our model system (to be introduced below) has found that
two scaling regions can be identified \cite{T2001}. While for small rates
the probability density function $\rho(\Gamma)$ scales as $\Gamma^{-1}$, at
larger scales it turns to $\Gamma^{-3/2}$ -- which is expected for strongly
transmitting channels from various models for transport through disordered
systems \cite{BGS1991}. Here we ask ourselves whether this prediction of a
smooth variation in the scaling of the monofractal behaviour (induced by
the smoothly changing exponent $\gamma$) can be generalized to characterize
the fluctuations on many scales using from the very beginning the technique
of multifractal analysis. Before we present our findings on the
multifractal scaling of the parametric fluctuations of the survival
probability, we introduce the kicked rotor system and our numerical
algorithm for the multifractal analysis in the subsequent two sections.

\section{Our transport model and the central observable}
\label{sec:Trans}

The $\delta-$kicked rotor is a widely studied, paradigmatic toy model of classical and
quantum dynamical theory \cite{Chi1979,Izr1990}. Using either cold or ultracold atomic gases,
the kicked rotor is realised experimentally by preparing a cloud of atoms with a
small spread of initial momenta, which is then subjected to a one-dimensional optical
lattice potential, flashed periodically in time \cite{MRBS1995}. In good approximation,
the Hamiltonian for the experimental realization of the rotor on the line
(in one spatial dimension) reads in dimensionless units \cite{MRBS1995}:
\begin{eqnarray}
\label{eq:ham}
\Hop(t^{\prime})
= \frac{p^2}{2} + k\,\cos{x}\;\sum_{t=1}^{\infty}\delta(t^{\prime}
 - t\,\tau)\;.
\end{eqnarray}
The derivation of the one-period quantum evolution operator exploits the spatial periodicity of
the potential by Bloch's theorem \cite{WGF2003}.
This defines quasimomentum $\beta$ as a constant of the motion, the value
of which is the fractional part of the physical momentum $p$ in dimensionless units
$p=n+\beta \;\;\; (n \in \mathbb{N})$.
Since $\beta$ is a conserved quantum number, $p$ can be labelled using its integer part $n$ only.
The spatial coordinate is then substituted by $\theta=x\;\mbox{mod}({2\pi})$ and
the quantum momentum operator by $\hat{\mathcal{N}}=-i\partial/\,\partial\theta$
with periodic boundary conditions.
The one-kick quantum propagation operator for a fixed $\beta$ is thus given by \cite{WGF2003}
\begin{eqnarray}
\label{eq:evol}
\hat{\mathcal{U}}_{\beta} = e^{-ik\,\cos(\hat{\theta})}\;
e^{-i\tau (\hat{\mathcal{N}}+\beta)^{2}/\;2} \;.
\end{eqnarray}

In close analogy to the transport problem across a solid-state sample,
we follow \cite{TMW2006,BCGT2001} to define the quantum survival probability as
the fraction of the atomic ensemble which stays within a specified region
of momenta while applying absorbing boundary conditions at the ``sample'' edges.
If we call $\psi(n)$ the wave function in momentum space and $n_{1}<n_{2}$ the
edges of the system, absorbing boundary conditions are implemented by setting
$\psi(n)\equiv 0$ if $n\leq n_{1} \equiv -1$ or $n\geq n_{2} \equiv 251$.
This truncation is carried out after each kick, and it mimics the escape of atoms
out of the spatial region where the dynamics induced by the Hamiltonian (\ref{eq:ham}) takes place.
If we denote by $\Pop$ the projection operator on the interval $]n_{1},n_{2}[$ the survival probability
after $t$ kicks is:
\begin{eqnarray}
P_{\rm sur}(t)= \left \| (\Pop \hat{\mathcal{U}}_{\beta}) ^{t} \psi(n,0) \right \|^{2} \;.
\end{eqnarray}

The early studies of the fluctuation properties of $P_{\rm sur}$ focused on its parametric
dependence on the quasimomentum $\beta$ \cite{T2001,BCGT2001}.
While $\beta$ is hard to control experimentally on a range of many
scales (with a typical uncertainty of 0.1 in experiments with an initial ensemble of
{\em ultra}-cold atoms \cite{Duffy2004}), some of us recently proposed to investigate
the parametric fluctuations as a function of the kicking period $\tau$
(see Eq.~(\ref{eq:ham})), which can be easily controlled on many scales in the experimental
realization of the model even with laser-cooled (just ``cold'') atoms \cite{TMW2006}.

In Figure 1 we present the survival probability $P_{\rm sur}$ of
the opened kicked rotor in the deep quantum regime (i.e., at
kicking periods $\tau \equiv \hbar_{\rm eff} > 1$ \cite{Izr1990})
as a function of the two different scan parameters $\beta$ and
$\tau$. The global oscillation with a period of the order 1 in
Fig.~1(a) originates from the $\beta$-dependent phase term
$\mathcal{N}\beta$ in the evolution operator (\ref{eq:evol}), and
can be understood qualitatively by remembering the Bloch band
structure of the corresponding quasienergy spectrum as a function
of $\beta$ \cite{Izr1990}. No such oscillating trend is found for
the graph as a function of the kicking period. Nonetheless, in the
following, we use a well developed variation of the standard
multifractal algorithm, which intrinsically takes account of such
global, yet irrelevant trends in the signal function $P_{\rm
sur}$. The basic feature of the {\em MultiFractal Detrended
Fluctuation Analysis} (MF-DFA) \cite{kantel} are now explained
before we present our central results which indicate the
multifractal scaling of data sets as the ones shown in Fig. 1.

\section{Multifractal detrended fluctuation analysis}
\label{sec:MFDA}

The MF-DFA is a generalization of the DFA method originally proposed by
\cite{peng}, and it is extensively described in \cite{kantel}. In the
recent years it was used, for instance, to investigate the nonlinear properties of
nonstationary series of wind speed records \cite{CSF24-05}, electro-cardiograms
\cite{mako}, and financial time series \cite{PHYA364-06}.

The method consists of five steps. First the series $\{x_i\}_{i=1}^N$ is
integrated to give the profile function:
\begin{equation}
y(k)=\sum_{i=1}^{k}(x_i -\bar{x})
\end{equation}
where $\bar{x}$ is the average value of $x_i$. The profile can be
considered as a random walk, which makes a jump to the right if $x_i
-\bar{x}$ is positive or to the left side  if $x_i -\bar{x}$ is negative.
In order to analyze the fluctuations, the profile is divided into
$N_s=int(N/s)$ non-overlapping segments of length $s$, and, since usually
$N$ is not an integer multiple of $s$, to avoid the cutting of the last
part of the series, the procedure is repeated backwards starting from the
end to the beginning of the data set. In each segment $\nu$ we subtract the
local polynomial trend of order $k$ and we compute the variance:
\begin{equation}
F^2(\nu,s)=\frac{1}{s}\sum_{i=1}^{s}
\left\{y[(\nu-1)s+i]-y_\nu^{k}(i)\right\}^2\;,
\end{equation}
for $\nu=1,\dots,N_s$, and
\begin{equation}
F^2(\nu,s)=\frac{1}{s}\sum_{i=1}^{s}
\left\{y[N-(\nu-N_s)s+i]-y_\nu^k(i)\right\}^2\;,
\end{equation}
with $\nu=N_s+1,\dots,2N_s$ for the backward direction. The order of the
polynomial defines the order of the MF-DFA too, therefore we may speak
about MF-DFA(1), MF-DFA(2), ... , MF-DFA($k$).

The fourth step consists on the averaging of all segments to
obtain the $q$-th order fluctuation function for segments of size
$s$:
\begin{equation}
F_q(s)=\left\{
\frac{1}{2N_s}\sum_{\nu=1}^{2N_s}[F^2(\nu,s)]^{q/2}\right\}^{1/q}.
\end{equation}

In the last step we determine the scaling behaviour of the fluctuation
function by analyzing the log-log plots of $F_q(s)$ versus $s$ for each
value of $q$. If the series is long-range correlated $F_q(s)$ increases for
large $s$ as a power law:
\begin{equation}
F_q(s)\sim s^{h(q)}.
\end{equation}
Since the number of segments becomes too small for very large scales
($s>N_s/4$), we usually exclude these scales for the fitting procedure to
determine $h(q)$. The MF-DFA reduces to the standard DFA for $q=2$, while
the scaling exponent $h(q)$ can be related to the standard multifractal
analysis considering stationary time series, in which $h(2)$ is identical
to the Hurst exponent $H$, therefore, $h(q)$ can be considered a
generalized Hurst exponent. Monofractal series indeed show a very weak or no
dependence of $h(q)$ on $q$. By example, for monofractal series as white
noise, the generalized Hurst exponent is $H=1/2$ for all $q$. On the
contrary, for multifractal time series, $h(q)$ is a function of $q$ and
this dependence influences the multifractality of the process. Referring to
the formalism of the partition function:
\begin{equation}
Z_q(s)=\sum_{\nu=1}^{N_s}|y_{\nu s}-y_{(\nu-1)s}|^q \sim s^{\tau(q)}
\end{equation}
where $\tau(q)$ is the Renyi exponent, to which the $h(q)$ is related by:
\begin{equation}
\tau(q)=1-qh(q).
\end{equation}
Now we are able to use the formalism of the multifractal spectrum
\cite{McCauley} $f(\alpha)$ to characterize the data set:
\begin{equation}
\label{eq:fa}
\begin{split}
&\alpha= \frac{d\tau(q)}{dq}=h(q)+q\frac{dh(q)}{dq} \\
&f(\alpha)=q\alpha - \tau(q)=q[\alpha -h(q)]+1 \;.
\end{split}
\end{equation}
The generalized dimensions are expressed as a function of
$\tau(q)$ or $h(q)$ \cite{JTAM35-05}:
\begin{equation}
D_q=\frac{\tau(q)}{q-1}=\frac{qh(q)-1}{q-1}\;,
\end{equation}
which cannot be not straightforwardly defined for $q=0$ and $q=1$.

If the signal is multifractal, the spectrum $f(\alpha)$ has approximately the form of an inverted parabola.
As significative parameters for its characterization we considered the point $\alpha_M$ corresponding to the maximum of $f(\alpha)$, and its
width $W_\alpha$ considered for a fixed $q$ interval. In other words,  $\alpha_M$ represents
the $\alpha$ value at which is situated
the ``statistically most significant part'' of the time series (i.e, the
subsets with maximum fractal dimension among all subsets of the series).
The width $W_\alpha$ is related to the dependence on $h(q)$
from q. The stronger this dependence, the wider is the fractal spectrum
(cf., eq. (\ref{eq:fa})).

\section{Results}\label{sec:res}
We performed a MF-DFA of order $k=1$ on data sets produced by
scanning the $\beta$ or $\tau$ parameter, respectively, over $10^5$ data points,
and considering different interaction times from $t=250$ to $t=10000$ kicks.
The analysis performed with higher order ($k=2$ and 3) polynomial detrending
for some of the series produced basically the same results. Furthermore, we tested our numerical algorithm on a
monofractal time series (white noise) and a well known multifractal process (binomial multifractal model \cite{Feder}).
For these two test series we reproduce the known analytical resuls, with a precision better than $1\%$.

A full analysis for $t=6000$ (see Figure
1) is shown in Figures 2 and 3 for
the $\beta$ and $\tau$ scanned series, respectively. Tables \ref{tab:beta} and
\ref{tab:tau} collect the multifractal parameters $\alpha_M$ and
$W_\alpha$, which were computed for $t=250\ldots 10000$. Analogously to \cite{EPL68-04},
we defined $W_\alpha$ as the width of the parabolic form of $f(\alpha)$ between
the points corresponding to $q=-3$ and $q=3$.

Figure 2(a) shows the scaling behaviour of the fluctuation
function $F_q(s)$, with $q\in[-5,5]$. Here $s$ represents the index of the
scanning parameter $\beta$, while the fit was performed in the zone
$\log(s)\in[1.6,2.7]$ (corresponding to $s\in[40,500]$). In Figure
2(b) we report the dependence of $h(q)$ on $q$, revealing the
multifractal nature of the data set. In order to better characterize the
multifractality and to highlight how it changes among the different
analyzed series, we have computed the MF spectrum $f(\alpha)$ (c.f.
2(c)). Figure 2(d) shows the variation of the
multifractal parameters for the different interaction times considered.
After a fast decrease, both the parameters tend to converge around the
values $\alpha_M=1.29$ and $W_\alpha=0.2$ (see also Tab. \ref{tab:beta}).
Very similar results were obtained for the $\tau$ scanned series (cf., Figure 3 and Table \ref{tab:tau}).
Comparing the values of Tables \ref{tab:beta} and \ref{tab:tau} we can say
that both the $\tau$ and the $\beta$ scanned series have essentially the
same multifractality.

Even if we cannot a priori predict the asymptotic similarity between the
two series of $\tau$ and $\beta$, we can a posteriori interpret this
result: both parameters enter not equally yet similarly in the {\em phase}
of the second factor on the right of eq. (2). As a consequence, the
restriction of $\beta$ to the unit interval does make no difference to the,
in principle, unboundedness of $\tau$ (in fact, to avoid different
dynamical properties of the system, $\tau$ was chosen in a restricted
window too, c.f. \cite{TMW2006}).

In general, two types of multifractality can be distinguished, and both of
them require different scaling exponents for small and large fluctuations.
(I) The multifractality can be due to the broad probability density
function for the values, and (II) it can also be due to different long
range correlations for small and large fluctuations. The simplest way to
distinguish between the mentioned two cases is to perform the analysis on a
randomly reshuffled series. The shuffling destroys all the correlations,
and the series with multifractals of type (II) will exhibit a monofractal
behaviour with $h_{\mbox{{\tiny shuf}}}(q)=0.5$ and $W_\alpha=0$. On the
contrary, multifractality of type (I) is not affected by the shuffling
procedure. If both (I) and (II) are present the series will show a weaker
multifractality than the original one.

We applied the shuffling procedure to the series showed in Figure
1. The procedure destroyed the multifractality of both
series since for both the sequences we obtained $h(q)=0.51\pm 0.01$ for $q\in[-5,5]$.
The dependence on $q$ was so weak that we were not able to compute any reliable
$f(\alpha)$ spectrum, which, in this case, can be considered singular, i.e., with
$W_\alpha \approx 0 $.

\section{Conclusions}
\label{sec:conc}

We studied the quantum kicked rotor, a paradigmatic model of quantum chaos,
which describes the time evolution of cold atoms in periodically flashed
optical lattices. Imposing absorbing boundary conditions allows one to
probe the transport properties of the system, here expressed by the
survival probability on a finite region in momentum space. For a fixed
interaction time, the quantum survival probability depends sensitively on
the parameters of the system, and our application of the detrended
multifractal method shows that {\em clear signatures of a multifractal
scaling of the survival probability} are found, as either the kicking
period or quasimomentum is scanned. Our results generalize the previously
predicted {\em mono}-fractal structure of the signal
\cite{GT2001,TMW2006,BCGT2001}, by characterizing long-range correlations
in the parametric fluctuations. In agreement with the monotonic increase of
the box counting dimension with the interaction time $t$ and its saturation
after $t \geqsi 5000$ observed in \cite{TMW2006}, we found a systematically
decreasing value for the maximum $\alpha_M$ of the MF spectrum and of its
widths $W_M$. Both of these two values also tend to saturate for $t \geqsi
5000$.

Future work along the lines of \cite{TMW2006} will be devoted to check in detail
whether traces of the here predicted multifractality could be
observed under real-life experimental conditions (e.g., for short interaction
times and finite resolutions in the scanning parameter \cite{TMW2006}).

\section{Acknowledgments}
S.W. acknowledges support by the Alexander von Humboldt Foundation
(Feodor-Lynen Program) and is grateful to Carlos Viviescas and Andreas
Buchleitner for their hospitality at the Max Planck Institute for the
Physics of Complex Systems (Dresden) where part of this work has been done. A.F. is
grateful to Holger Kantz and Nikolay Vitanov for their support and
important suggestions. Furthermore we thank Riccardo Mannella for his helpful
advice on the numerical procedure.

\begin{figure}
  \centering
  \includegraphics[width=\linewidth, keepaspectratio=true]{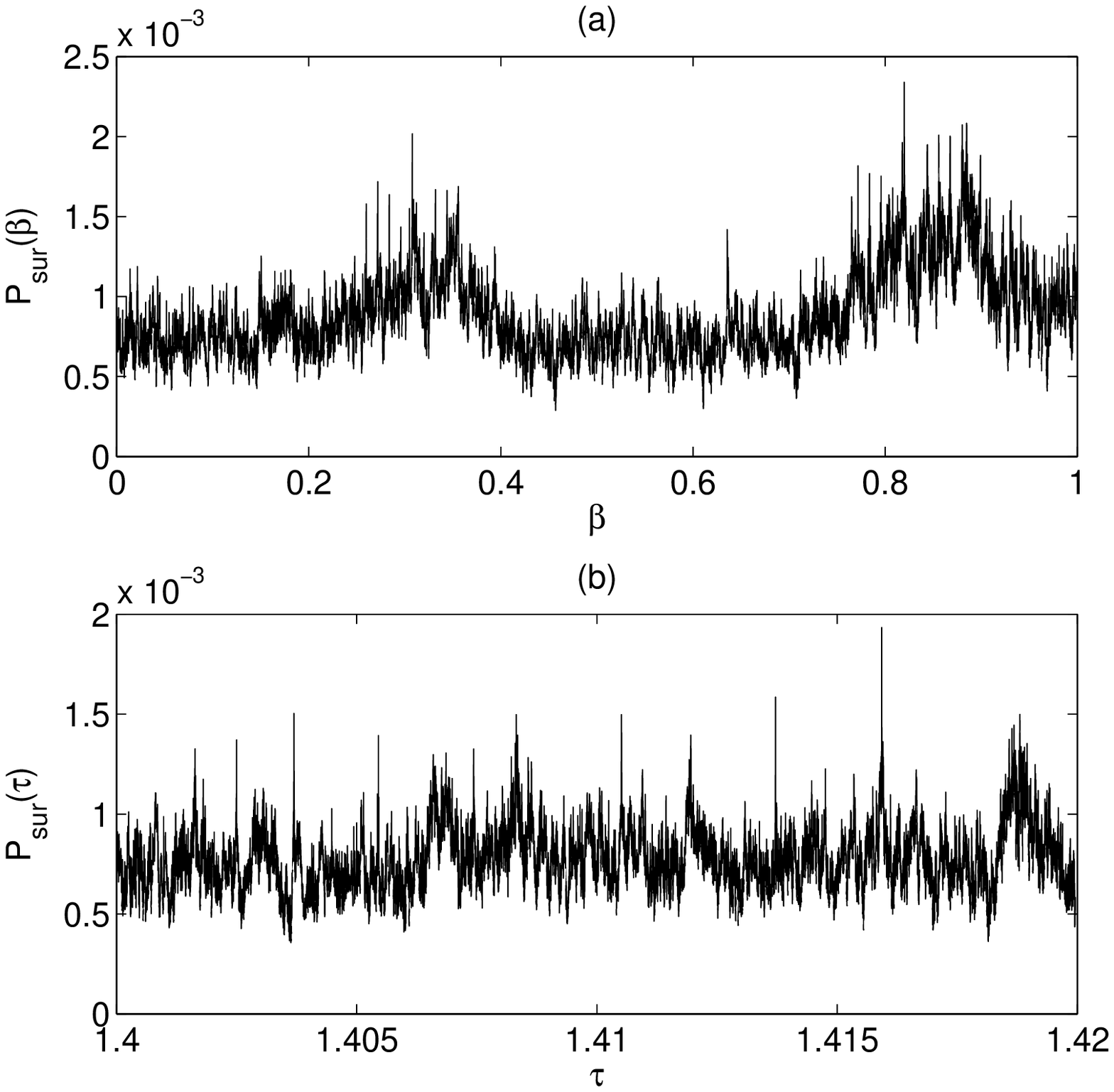}
  \caption{ The series $P_{sur}(\beta)$ (a) and $P_{sur}(\tau)$ (b) after an interaction time
            of $t=6000$ kicks. Other paramters are $k=5$, and $\tau = 1.4$ in (a) and $\beta = 0$ in (b),
            respectively. Both sequences extend over $10^5$ sampling points along the
            shown intervals.}
  \label{fig:serie}
\end{figure}

\begin{figure}
  \centering
\includegraphics[width=\linewidth, keepaspectratio=true]{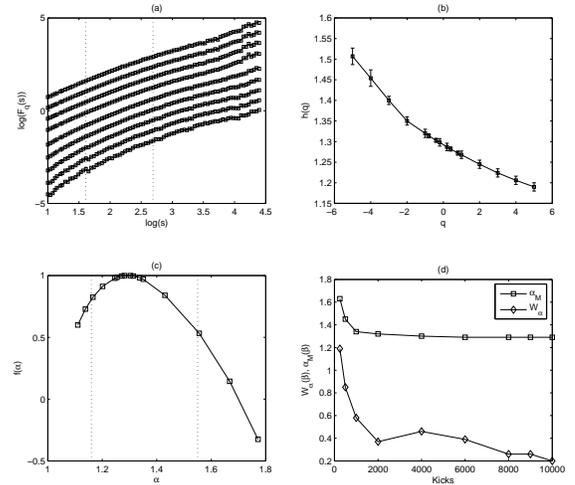}
\caption{(a) The decimal logarithm of the fluctuation function
$F_q(s)$ for $q\in[-5,5]$ for the $\beta$ scanned series after an
interaction time of 6000 kicks. The fitting procedure was
performed in the zone $\log(s)\in[1.6,2.7]$ (corresponding to
$s\in[40,500]$). The curves were vertically shifted for better
reading. (b) The spectrum of the generalized Hurst exponents
$h(q)$; its strong dependence on $q$ indicates the multifractal
behaviour. The error bars show the uncertainty arising from the
fits to the curves in (a). (c) The $f(\alpha)$ spectrum with
$\alpha_M=1.29$ and $W_\alpha = 0.39$. The dotted lines indicate
the $\alpha$ interval used to compute $W_\alpha$. (d) The
multifractal parameters $\alpha_M$ and $W_\alpha$ as a function of
the interaction time. After a strong, initial variation, the value
$\alpha_M$ shows a saturation towards the value $\alpha_M \approx
1.3$. The width $W_\alpha$ shows approximately the same behaviour,
and tends to saturate towards the value $W_\alpha \approx 0.2$.}
\label{fig:beta}
\end{figure}

\begin{figure}
\centering
\includegraphics[width=\linewidth, keepaspectratio=true]{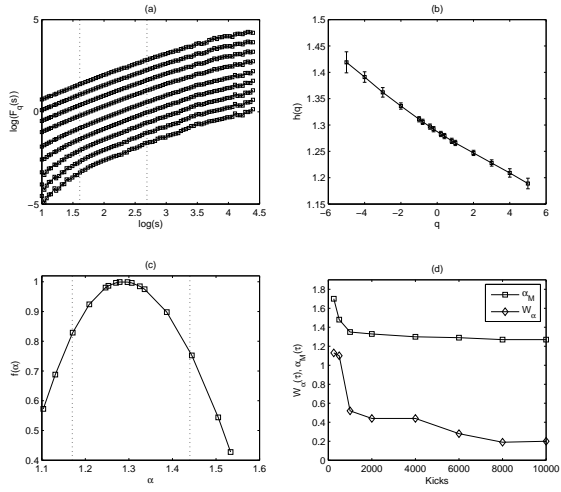}
\caption{(a) The decimal logarithm of $F_q(s)$ with $q\in[-5,5]$
for the $\tau$ scanned series with $t=6000$. The fitting procedure
was performed in the zone $\log(s)\in[1.6,2.7]$ (corresponding to
$s\in[40,500]$). (b) The spectrum of the generalized Hurst
exponents $h(q)$. (c) The $f(\alpha)$ spectrum with $\alpha_M =
1.29$ and $W_\alpha = 0.28$. (d) The variation of the multifractal
parameters $\alpha_M$ and $W_\alpha$ with the interaction time.
$\alpha_M$ shows a saturation towards the value $\alpha_M \approx
1.3$, while $W_\alpha$ tends to saturate around the value
$W_\alpha \approx 0.2$.}
  \label{fig:tau}
\end{figure}


%
%
%
%

\begin{center}
\begin{table}
  \caption{Multifractal parameters of the $\beta$-scanned data sets for different interaction times.
           The estimated error due to the fitting procedure described in section \ref{sec:MFDA} is about
           $\pm 0.02$ for $\alpha_M$ and $\pm 0.05$ for $W_{\alpha}$.}
  \label{tab:beta}
  \begin{tabular}{ccc}
    \hline
           & $\alpha_M$ & $W_{\alpha}$\\
    \hline
    t=250    & 1.63 & 1.19 \\
    t=500    & 1.45 & 0.85 \\
    t=1000   & 1.34 & 0.58 \\
    t=2000   & 1.32 & 0.37 \\
    t=4000   & 1.30 & 0.46 \\
    t=6000   & 1.29 & 0.39 \\
    t=8000   & 1.29 & 0.26 \\
    t=10000  & 1.29 & 0.20 \\
     \hline
  \end{tabular}
\end{table}
\end{center}

\begin{center}
\begin{table}
 \caption{Multifractal parameters of the $\tau$-scanned data sets for different interaction times
          (error estimates as stated for the preceding table).
           }
  \label{tab:tau}
  \begin{tabular}{ccc}
    \hline
           & $\alpha_M$ & $W_{\alpha}$\\
    \hline
    t=250    & 1.70 & 1.13 \\
    t=500    & 1.48 & 1.12 \\
    t=1000   & 1.35 & 0.9 \\
    t=2000   & 1.33 & 0.52 \\
    t=4000   & 1.30 & 0.44\\
    t=6000   & 1.29 & 0.28 \\
    t=8000   & 1.27 & 0.19 \\
    t=10000  & 1.27 & 0.20\\
     \hline
  \end{tabular}

\end{table}
\end{center}


\end{document}